\documentclass[prb,preprint]{revtex4} 

\usepackage{amsmath}  
\usepackage{amsfonts} 
\usepackage{graphicx} 
\newcommand{\gothg}{\mathfrak{g}}

\begin{document}


\title{The Schwarzschild metric: It's the coordinates, stupid!}

\author{Pierre Fromholz}
\email{fromholz@clipper.ens.fr} 
\affiliation{\'Ecole Normale Sup\'erieure, 75005 Paris, France } 

\author{Eric Poisson}
\email{epoisson@uoguelph.ca}
\affiliation{Department of Physics, University of Guelph, Guelph ON N1G 2W1 Canada}

\author{Clifford M. Will}
\email{cmw@physics.ufl.edu}
\affiliation{Department of Physics, University of Florida, Gainesville FL 32611\\
Institut d'Astrophysique, 75014 Paris, France}

\date{\today}

\begin{abstract}
Every general relativity textbook emphasizes that coordinates have no physical meaning.  Nevertheless, a coordinate choice must be made in order to carry out real calculations, and that choice can make the difference between a calculation that is simple and one that is a mess.   We give a concrete illustration of the maxim that ``coordinates matter'' using the exact Schwarzschild solution for a vacuum, static spherical spacetime.  We review the standard textbook derivation, Schwarzschild's original 1916 derivation, and a derivation using the Landau-Lifshitz formulation of the Einstein field equations.  The last derivation is much more complicated, has one aspect for which we have been unable to find a solution, and gives an explicit illustration of the fact that the Schwarzschild geometry can be described in infinitely many coordinate systems.
\end{abstract}

\maketitle 

\section{Introduction}
\label{sec1} 

Every student of general relativity is taught that coordinates are irrelevant to physics.  The principle of general covariance, upon which general relativity is built, implies that coordinates are simply labels of spacetime events that can be assigned completely arbitrarily (subject to some conditions of smoothness and differentiability).  The only quantities that have physical meaning -- the measurables -- are those that are invariant under coordinate transformations. 
One such invariant is the number of ticks on an atomic clock giving the proper time between two events.

Yet apart from some highly abstract mathematical topics, virtually everything covered in a typical course in general relativity uses coordinates.   The  reason is simple: it's hard to do explicit calculations -- derivatives, products, sums -- without coordinates.   It is here where coordinates matter\cite{carville}.  

Indeed the choice of coordinates can be critical when it comes to trying to solve Einstein's equations for situations of interest.  In linearized general relativity, for example, the use of coordinates defined by the Lorenz gauge leads to a simple wave equation.   

The choice of coordinates is critical even for the oldest and most famous exact solution of Einstein's equations, the Schwarzschild metric.   Schwarzschild's choice of coordinates was not the one used in standard textbook treatments, yet it did permit him to find the solution in a straightforward way, very shortly after the publication of Einstein's initial papers on general relativity\cite{schwarzschild,antoci99,zelmanov}.  On the other hand, his coordinate choice led to considerable confusion about the nature of what came to be known as the ``Schwarzschild singularity''.

Schwarzschild found his solution by solving the vacuum field equations in the form $R_{\mu\nu} =0$, where $R_{\mu\nu}$ is the Ricci tensor constructed from the metric.   
But there is a rather different formulation of Einstein's equations, known as the Landau-Lifshitz formulation\cite{LL}.   In this version of Einstein's equations, the basic variable is not the spacetime metric $g_{\mu\nu}$, but the ``gothic inverse metric'', $\gothg^{\alpha\beta} \equiv \sqrt{-g} g^{\alpha\beta}$, where $g$ is the determinant of $g_{\mu\nu}$.   After adopting a coordinate system specified by the four conditions
$\partial_\beta \gothg^{\alpha\beta} = 0 $,
called harmonic gauge, one can write the Einstein equations as a flat spacetime wave equation for $\gothg^{\alpha\beta}$, whose source consists of the energy-momentum tensor for matter, plus terms that, with one exception, are quadratic in first derivatives of $\gothg^{\alpha\beta}$.

During the past 35 years, this formulation of general relativity has been developed extensively as the basis for post-Newtonian theory, post-Minkowskian theory, and gravitational-wave physics.   This is because, in a weak-field approximation, one can write $\gothg^{\alpha\beta} = \eta^{\alpha\beta} - h^{\alpha\beta} $, where $\eta^{\alpha\beta}$ is a background Minkowski metric and where the field $h^{\alpha\beta}$ is ``small'' in a suitable sense.  Since the right-hand-side of the wave equation consists of matter terms or terms that (for the most part) are quadratic in derivatives of $h^{\alpha\beta}$, the equation can be iterated, leading to successively higher-order approximations for the field.  

This approach has found its fullest utility in the program of calculating the orbital evolution of inspiralling binaries of compact objects (black holes or neutron stars) and the gravitational waves emitted, to very high orders in a post-Newtonian (PN) expansion beyond the lowest-order Newtonian motion and the lowest-order quadrupole formula for radiation.  It is commonplace today to see equations of motion for such binaries displayed through 3.5 PN order ($O(v/c)^7$ beyond Newtonian gravity) and the gravitational waveform and the evolution of the orbital frequency displayed to similar orders beyond the leading terms\cite{lucLRR}. The effects of spin have also been incorporated to high post-Newtonian orders.   Post-Newtonian theory has proven to be ``unreasonably effective'' (in the words of Eugene Wigner), for example in accurately describing such inpirals well into a regime where the fields are not so weak and the orbital motions are not so slow, matching smoothly onto results from numerical relativity, which then describes the final inspiral and merger of the two compact bodies\cite{willpnas}.

Post-Minkowskian and post-Newtonian theories have become so central to gravitational-wave physics and astrophysics, that two of us (EP and CMW) have recently completed a textbook entitled {\em Gravity: Newtonian, post-Newtonian, Relativistic}\cite{poissonwill} that uses the Landau-Lifshitz formulation as the centerpiece of its discussion of general relativity.   Like any textbook, it features exercises at the end of each chapter, and thus, in writing the chapters in which the  Landau-Lifshitz formulation is laid out, we imagined posing the obvious exercise for the student:  Solve these equations exactly in vacuum for static, spherical symmetry, and thus obtain the Schwarzschild metric.  What could be simpler?

This turned out to be not at all simple, and the exercise provides the perfect illustration of the dictum that, while the choice of coordinates has no physical meaning whatsoever, it can have a big impact on the ease of finding a solution of Einstein's equations.   Furthermore, although one solution of the LL equations yields the Schwarzschild metric in the so-called harmonic radial coordinate $r_{\rm H}$, related to the standard Schwarzschild coordinate $r_{\rm S}$ by $r_{\rm H} = r_{\rm S} - M$, where $M$ is the mass of the object, there is an infinite set of additional solutions, for which we are unable to obtain closed form expressions.  Exploring the nature of these additional solutions yields insights into the nature of coordinate freedom that one does not get from standard textbook treatments of the Schwarzschild metric.  

We begin in Sec. \ref{sec2} by reviewing the ``textbook'' derivation of the Schwarzschild metric, which has been honed and refined so as to be as simple as possible.   We then turn in Sec.\ \ref{sec3} to Schwarzschild's original 1916 derivation, which, while very similar to the textbook version, has one key difference.  Section \ref{sec4} treats the Landau-Lifshitz version of the derivation.  Concluding remarks are made in Sec.\ \ref{sec5}.   We use units in which $G = c =1$; Greek indices span the four spacetime dimensions, while Latin indices span the three spatial dimensions, and we adopt the standard Einstein summation convention on repeated indices;  partial derivatives are denoted by $\partial_\alpha$.

\section{The ``textbook'' solution}
\label{sec2}

The standard textbook solution begins by exploiting the spherical symmetry of the problem to show that the metric can always be written in the form
\begin{equation}
ds^2 = - e^{2\Phi(r_{\rm S})} dt^2 + \frac{dr_{\rm S}^2}{1-2m(r_{\rm S})/r_{\rm S}} + r_{\rm S}^2 \left ( d\theta^2 + \sin^2 \theta d\phi^2 \right ) \,,
\label{eq:metricansatz}
\end{equation}
Here $\theta$ and $\phi$ are the standard coordinates of the two-sphere, $r_{\rm S}$ has the interpretion of being ${\cal C}/2\pi$, where $\cal C$ is the physically measured circumference of a surface of constant $r_{\rm S}$, $t$ and $\phi$, or $\sqrt{{\cal A}/4\pi}$, where $\cal A$ is the physically measured area of a surface of constant $r_{\rm S}$ and $t$.  It is often called the Schwarzschild radial coordinate.  The functions $\Phi(r_{\rm S})$ and $m(r_{\rm S})$ are arbitrary functions of the radial coordinate; to be fair, the form of $g_{r_{\rm S}r_{\rm S}}$ has been chosen with considerable hindsight, yet it is still an arbitrary function of $r_{\rm S}$.  We assume for simplicity that the metric is static (the form shown in Eq.\ (\ref{eq:metricansatz}) turns out to be also valid for a dynamical spherically symmetric spacetime, by allowing $\Phi$ and $m$ to be functions of time $t$ as well as of $r_{\rm S}$).   

Given the metric, it is relatively straightforward to calculate the Christoffel symbols, the Riemann and Ricci tensors, the Ricci scalar, and the Einstein tensor.  These days, in fact, one obtains these tensors using well-known software packages.  The vacuum Einstein equations are most simply expressed as $G^\mu_\nu = 0$, of which the only non-trivial equations are
\begin{eqnarray}
G^0_0 &=& -\frac{2}{r_{\rm S}^2} \frac{ dm(r_{\rm S})}{dr_{\rm S}} = 0 \,, 
\nonumber \\
G^{r_{\rm S}}_{r_{\rm S}} &=& - \frac{2}{r_{\rm S}} \left ( 1- \frac{2m(r_{\rm S})}{r_{\rm S}} \right ) \frac{ d\Phi(r_{\rm S})}{dr_{\rm S}}  - \frac{2m(r_{\rm S})}{r_{\rm S}^3} = 0 \,.
\end{eqnarray}
The solutions are immediate, and after imposing the boundary condition that $\Phi(r_{\rm S}) \to 0$ as $r_{\rm S} \to \infty$, one finds
\begin{eqnarray}
m(r_{\rm S}) &=& M = \, {\rm constant} \,,
\nonumber \\
\Phi(r_{\rm S}) &=& -\frac{1}{2} \ln | 1 - 2M/r_{\rm S} | \,,
\end{eqnarray}
yielding the canonical form of the vacuum Schwarzschild metric,\begin{equation}
ds^2 = -(1-2M/r_{\rm S})dt^2 + \frac{dr_{\rm S}^2}{1-2M/r_{\rm S}} + r_{\rm S}^2
\left ( d\theta^2 + \sin^2 \theta d\phi^2 \right ) \,.
\label{Schmetric}
\end{equation}
By examining the orbits of test bodies far from the center, one identifies $M$ as the so-called Kepler-measured mass of the body.  Under the stated conditions, the solution is unique, up to the parametrization by the mass $M$.

This solution is so iconic that few textbooks even mention the existence of other radial coordinate systems, such as the {\em isotropic} coordinate $r_{\rm iso}$, related to the Schwarzschild coordinate  $r_{\rm S}$ by 
$r_{\rm S} = r_{\rm iso} (1+ M/2r_{\rm iso})^2$, in which the spatial part of the metric is proportional to the Euclidean metric, or the {\em harmonic} coordinate
$r_{\rm H} = r_{\rm S} - M$ (more on $r_{\rm H}$ later).

\section{Schwarzschild's solution of 1916}
\label{sec3}

In the fall of 1915, Karl Schwarzschild was already a well-known German astronomer, director of the Astrophysical Observatory in Potsdam and a member of the Prussian Academy of Sciences.  At the outbreak of World War I, he volunteered for service despite being over 40 years of age\cite{remark}, and served in France and Russia.  But on the eastern front, he contracted a rare auto-immune skin disease called pemphigus.  While confined to hospital, he attempted to find exact solutions of Einstein's equations of general relativity, newly published in November 1915.  He obtained the solution for both a spherically symmetric star of uniform density and of a ``mass point''.  The latter solution will concern us here.  His results were published in early 1916\cite{schwarzschild},  but Schwarzschild soon died of the disease, in May, 1916\cite{myth}.

Unfortunately, those November papers of Einstein were infected with Einstein's obsession with coordinate systems in which the determinant of the metric was precisely $-1$.  In part because such systems made calculations of tensorial quantities such as the Ricci tensor simple, Einstein insisted that these were somehow physically privileged coordinate systems.  By the time he published the fully developed theory of general relativity in May of 1916, he had given up this notion and fully embraced the concept of general covariance, the idea that any coordinate system is allowed and that coordinates have no physical significance.

But based on the November 1915 papers that were available to him, Schwarzschild was forced to find a solution using coordinates in which $g \equiv \det (g_{\mu\nu})= -1$.   This was awkward, because even in flat spacetime in spherical coordinates, $g = -\rho^4 \sin^2 \theta$, where we will use $\rho$ provisionally to denote the radial coordinate.   Schwarzschild got around this by defining a new radial coordinate $x = \rho^3/3$ and a new angular coordinate $\psi = - \cos \theta$.
In these coordinates, the metric of flat spacetime has the form
\begin{equation}
ds^2 = -dt^2 + \frac{dx^2}{\rho^4} + \rho^2 \left ( \frac{d\psi^2}{\sin^2 \theta} + \sin^2 \theta d\phi^2 \right ) \,,
\end{equation}
for which $g=g_{00} g_{xx} g_{\psi\psi} g_{\phi\phi} =-1$.   The $(t, x, \psi, \phi)$ coordinates are thus a privileged set of coordinates, in Einstein's view.

To solve Einstein's equations in static spherical symmetry, Schwarzschild then proposed the metric
\begin{equation}
ds^2 = -f_0 dt^2 + f_1 dx^2 + f_2 \left ( \frac{d\psi^2}{\sin^2 \theta} + \sin^2 \theta d\phi^2 \right ) \,,
\label{Schansatz}
\end{equation}
where $f_0$, $f_1$ and $f_2$ are functions only of $x$,
along with the requirement that $f_0 f_1 f_2^2 = 1$.  The asymptotically flat boundary conditions he imposed were $f_0 \to 1$, $f_1 \to \rho^{-4}$ and $f_2 \to \rho^2$ as $x \to \infty$.  

Given the form of the metric it is again straightforward, either by hand or by software, to calculate the tensors needed for Einstein's equations.  Interestingly, the condition $g=-1$ makes the vacuum field equation $R_{\mu\nu} =0$ rather simple in the privileged coordinates, namely
\begin{equation}
\partial_\alpha \Gamma^\alpha_{\mu\nu} + \Gamma^\alpha_{\mu\beta}
\Gamma^\beta_{\nu\alpha} = 0 \,,
\end{equation}
where $\Gamma^\alpha_{\mu\nu}$ are the Christoffel symbols.

As before, only the $(00)$ and $(xx)$ components of Einstein's equations are needed, together with the condition $f_0f_1f_2^2 =1$, and the solution can be found by simple means.  The result is
\begin{eqnarray}
f_0 &=& 1 - \frac{2M}{(3x+ b)^{1/3}} \,,
\nonumber \\
f_1 &=& \frac{(3x+b)^{4/3}}{1-2M/(3x+ b)^{1/3}} \,,
\nonumber \\
f_2 &=& (3x+ b)^{2/3} \,,
\end{eqnarray}
where one constant of integration has been fixed to be $2M$ by examining the metric far from the source; $b$ is a second constant of integration.  Schwarzschild noticed that by defining a new variable
\begin{equation}
r_{\rm S} \equiv (3x+ b)^{1/3} = (\rho^3 + b)^{1/3} \,,
\end{equation}
he could put the metric (\ref{Schansatz}) into a simpler form, which is precisely Eq.\ (\ref{Schmetric}).

But Schwarzschild went on to address the integration constant $b$.  He demanded that the metric be regular everywhere except at the location of the mass-point, which he assigned to be at $\rho = 0$, where the metric should be singular.  This fixed $b = (2M)^3$.  This choice resulted in considerable confusion about the nature of the ``Schwarzschild singularity'', which was not cleared up fully until the 1960s\cite{israel}.  Because we now are attuned to the complete arbitrariness of coordinates, we understand that $\rho = 0$, or $r_{\rm S} = 2M$ is not the origin, but is the location of the event horizon, while $\rho = -2M$, or $r_{\rm S} = 0$ is the location of the true physical singularity inside the black hole\cite{note}.  

The unusual radial coordinate $x$ was forced on Schwarzschild by Einstein's constraint $g=-1$, nevertheless it led to a quite simple derivation of the exact solution.   In the next section, we will encounter a very different approach that yields a much more complicated set of equations and an interesting surprise. 

\section{Solution using the Landau-Lifshitz formulation}
\label{sec4}

Einstein's equations can be formulated in an alternative manner, sometimes called the Landau-Lifshitz (LL) approach, or the ``relaxed'' Einstein equations.  The basic variable is not the metric $g_{\mu\nu}$, but the ``gothic inverse metric'', 
\begin{equation}
\mathfrak{g}^{\alpha\beta} \equiv \sqrt{-g} g^{\alpha\beta} \,.
\end{equation}
One next adopts a coordinate system specified by the four conditions
\begin{equation}
\partial_\beta \mathfrak{g}^{\alpha\beta} = 0 \,.
\label{Lorenz}
\end{equation}
We will refer to this coordinate system as the harmonic gauge (it is also known as deDonder gauge).  Einstein's equations can then be written in the form
\begin{equation} 
\Box \mathfrak{g}^{\alpha\beta} = 16\pi (-g)\bigl( T^{\alpha\beta}+ t^{\alpha\beta}_{\rm LL} + t^{\alpha\beta}_{\rm H} \bigr)  \,,
\label{wave_equation}
\end{equation}
where $\Box \equiv \eta^{\alpha\beta} \partial_\alpha \partial_\beta = -\partial_t^2 + \nabla^2$ is the flat-spacetime d'Alembertian, with $\eta^{\alpha\beta} = {\rm diag} (-1,1,1,1)$ the Minkowski metric;
$T^{\alpha\beta}$ is the energy-momentum tensor of matter;
$t^{\alpha\beta}_{\rm LL}$ is the Landau-Lifshitz pseudotensor, given, after imposing the gauge condition, by
\begin{eqnarray} 
16\pi(-g) t^{\alpha\beta}_{\rm LL} &=& 
\frac{1}{2} g^{\alpha\beta} g_{\lambda\mu} 
  \partial_\rho \gothg^{\lambda\nu} 
  \partial_\nu \gothg^{\mu\rho} 
- g^{\alpha\lambda} g_{\mu\nu} 
  \partial_\rho \gothg^{\beta\nu} 
  \partial_\lambda \gothg^{\mu\rho} 
  \nonumber \\ & & 
- g^{\beta\lambda} g_{\mu\nu} 
  \partial_\rho \gothg^{\alpha\nu} 
  \partial_\lambda \gothg^{\mu\rho} 
+ g_{\lambda\mu} g^{\nu\rho} 
  \partial_\nu \gothg^{\alpha\lambda} 
  \partial_\rho \gothg^{\beta\mu} 
\nonumber \\ & &
+ \frac{1}{8} \bigl( 2 g^{\alpha\lambda} g^{\beta\mu} 
  - g^{\alpha\beta} g^{\lambda\mu} \bigr) 
   \bigl( 2 g_{\nu\rho} g_{\sigma\tau} - g_{\rho\sigma} g_{\nu\tau}
  \bigr) \partial_\lambda \gothg^{\nu\tau} 
  \partial_\mu \gothg^{\rho\sigma}  \,,
\label{LL_pseudotensor} 
\end{eqnarray} 
and 
$t^{\alpha\beta}_{\rm H}$ is an additional pseudotensor related to the imposition of harmonic gauge, given by
\begin{equation} 
16\pi (-g) t^{\alpha\beta}_{\rm H} = 
\partial_\mu \gothg^{\alpha\nu} \partial_\nu \gothg^{\beta\mu}
- (\gothg^{\mu\nu} - \eta^{\mu\nu}) \partial_{\mu\nu} \gothg^{\alpha\beta} \,.
\label{harmonic_pseudotensor} 
\end{equation}
Equation (\ref{wave_equation}) is exact, and is completely equivalent to Einstein's equations, $G_{\mu\nu} = 8\pi T_{\mu\nu}$.

The first step is to write down a form for the metric.  The LL formulation is defined to work in Cartesian-like coordinates, in which $\eta^{\alpha\beta} = {\rm diag}(-1,1,1,1)$.    We define the spatial vector $x^i$, the Cartesian metric $\delta_{ij}$, with $r^2 \equiv  \delta_{ij} x^i x^j$, and the unit vector $n^i \equiv x^i/r$.  We write down a general form for the gothic inverse metric $\gothg^{\alpha\beta}$ appropriate for static spherical symmetry:
\begin{eqnarray}
\gothg^{00} &=& N(r) \,,
\nonumber \\
\gothg^{0j} &=& 0 \,,
\nonumber \\
\gothg^{jk} &=& \alpha(r) P^{jk} + \beta(r) n^j n^k \,,
\label{eq:gothgform}
\end{eqnarray}
where $P^{jk} := \delta^{jk} - n^j n^k$ projects to the subspace orthogonal to $n^j$.  (The student is asked to justify this form.)  The choice of 
$P^{jk}$ and $n^jn^k$ as the two spatial tensors instead of $\delta^{jk}$ and $n^jn^k$ is made purely for convenience;  with this choice, the inverse $\gothg_{\alpha\beta}$ has the same form as Eq.\ (\ref{eq:gothgform}), but with the coeffients $1/N$, $1/\alpha$ and $1/\beta$, respectively.  

The harmonic gauge condition $\partial_\beta \gothg^{\alpha\beta} = 0$ leads to the constraint
\begin{equation}
\beta' = \frac{2}{r} (\alpha- \beta) \,,
\label{eq:gauge}
\end{equation}
where ``prime'' denotes a radial derivative, $d/dr = n^j \partial_j$.
Just as before, then, there are two functions to be determined from Einstein's equations.   We must solve those equations subject to the boundary conditions $N(r) \to -1$, $\alpha(r) \to 1$ and $\beta(r) \to 1$ as $r \to \infty$.

The left-hand side of the relaxed equations become $\Box \gothg^{00} = \nabla^2 N$ and $\Box \gothg^{jk} = \nabla^2 \gothg^{jk}$.  On the right-hand side the matter energy-momentum tensor $T^{\alpha\beta}$ vanishes, and the Landau-Lifshitz and harmonic pseudotensors have the form
\begin{eqnarray}
16\pi (-g) t_{\rm LL}^{00}
&=& N\left [ \frac{7}{8} \beta \frac{{N'}^2}{N^2}  + \frac{3}{8} \frac{{\beta'}^2}{\beta}   - \frac{1}{2} \frac{\alpha' \beta'}{\alpha} 
+ \frac{1}{2} \beta \frac{N'\alpha'}{N\alpha} + \frac{1}{4} \frac{N'\beta'}{N}
\right ] \,,
\nonumber \\
16\pi (-g) t_{\rm LL}^{jk}
&=& P^{jk} \left [ \frac{3}{8} \alpha \frac{\beta'^2}{\beta} +\beta \frac{\alpha'^2}{\alpha} - \frac{1}{8} \alpha \beta \frac{N'^2}{N^2} + \frac{1}{2} \beta \frac{N' \alpha'}{N} +\frac{1}{4} \alpha \frac{N' \beta'}{N} + \frac{1}{2} \alpha' \beta' \right]
\nonumber \\ && \quad
+ n^j n^k \left [
\frac{1}{8}\beta'^2 + \frac{1}{2} \beta \frac{\alpha' \beta'}{\alpha} +\frac{1}{8} \beta^2 \frac{N'^2}{N^2} - \frac{1}{2} \beta^2 \frac{N'\alpha'}{N\alpha} - \frac{1}{4} \beta \frac{N'\beta'}{N}
\right ] \,,
\nonumber \\
16\pi (-g) t_{\rm H}^{00}
&=&
 \nabla^2 N - 2\alpha \frac{N'}{r} - \beta N'' \,,
\nonumber \\
16\pi (-g) t_{\rm H}^{jk}
&=& \nabla^2 \mathfrak{g}^{jk} - P^{jk} \left [ \beta \alpha'' + 2\alpha\frac{\alpha'}{r} + \alpha ' \beta' 
- \alpha \frac{\beta'}{r} 
\right ]
\nonumber \\ && \quad
+ n^jn^k \left [ \beta'^2 + 2 \beta \frac{\beta' - \alpha'}{r} - 3 \alpha \frac{\beta'}{r}
\right ] \,.
\end{eqnarray}
Note that $\nabla^2 N$ and $\nabla^2 \gothg^{jk}$ actually cancel between the left-hand and right-hand sides of the field equations, while $N''$ and $\alpha''$ appear elsewhere in the harmonic pseudotensor.  The gauge condition (\ref{eq:gauge}) has been used liberally to simplify the equations, whose initial forms are much messier than this.  
Putting together the field equations, equating separately the coefficients of $P^{jk}$ and $n^jn^k$ in the $(jk)$ components, and defining the new variables
\begin{equation}
X \equiv \frac{\alpha'}{\alpha} \,, \quad 
Y \equiv \frac{\beta'}{\beta} \,, \quad 
Z \equiv \frac{N'}{N} \,,
\end{equation}
we can put the field equations in the form
\begin{subequations}
\begin{eqnarray}
X' + XY + \frac{1}{r} (2X-Y) &=& Q \,,
 \label{XYZa}\\
XY + \frac{1}{r} (2X+Y) &=& -Q \,,
 \label{XYZb} \\
Z' + YZ + \frac{2}{r} Z &=& Q \,,
 \label{XYZc}
\end{eqnarray}
 \label{XYZabc}
\end{subequations}
where 
\begin{equation}
Q \equiv \frac{1}{8} \Bigl( 3Y^2 - Z^2 + 2 YZ + 4XZ - 4XY \Bigr) \,.
\end{equation}
Combining Eqs.\ (\ref{XYZa}) and (\ref{XYZb}) leads immediately to 
$X'/X + 2Y + 4/r = 0$, which integrates to $r^4 \beta^2 \alpha'/\alpha = C$, where $C$ is an integration constant.  

\subsection{The case $C=0$}

Setting $C=0$ implies $\alpha = 1$ after imposing the boundary condition at infinity.   It is then straightforward to solve the gauge condition (\ref{eq:gauge}) for $\beta$, with an additional integration constant $D$, and then to substitute the results into (\ref{XYZb}) and to solve for $N$.  One can verify that Eq.\ (\ref{XYZc}) is then satisfied automatically.  The constant $D$ is linked to the mass $M$ by looking at the Newtonian limit, with the final result that 
\begin{eqnarray}
N &=& - \frac{(1+M/r)^3}{1-M/r}  \,,
\nonumber \\
\alpha &=&1 \,,
\nonumber \\
\beta &=& 1- \left ( \frac{M}{r} \right )^2 \,.
\label{Nba}
\end{eqnarray}
Obtaining the metric from $\gothg^{\alpha\beta}$, we find the final form
\begin{equation}
ds^2 = - \left ( \frac{1-M/r}{1+M/r} \right ) dt^2 + 
\left ( \frac{1+M/r}{1-M/r} \right ) dr^2 + (r+M)^2 \left ( d\theta^2 + \sin^2 \theta d\phi^2 \right ) \,.
\label{SchmetricHarmonic}
\end{equation}
This is completely equivalent to the Schwarzschild metric, as can be shown  by the simple transformation 
\begin{equation}
r = r_{\rm S} - M \,.
\label{eq:radialtransform}
\end{equation}
In  Eq.\ (\ref{SchmetricHarmonic}), $r$ is the Harmonic radial coordinate $r_{\rm H}$.

\subsection{The case $C \ne 0$}

By combining the equation $r^4 \beta^2 \alpha'/\alpha = C$ with the gauge condition (\ref{eq:gauge}), we can obtain a second-order differential equation for $\beta$ alone given by
\begin{equation}
W'' - \frac{W'}{r} = C \frac{W'}{W^2}  \,,
\label{eq:W}
\end{equation}
where $W \equiv r^2 \beta$.   The trivial solution $W' = 0$ implies that $\beta \propto 1/r^2$, which violates the boundary condition $\beta \to 1$ at infinity.  
We have been utterly unable to find a closed form or analytic solution to this non-linear differential equation. We have tried a wide range of changes of variables.  For example, setting $W(r)=e^{\rho/2}H(\rho)$ with $\rho= \ln(r)$, we get
\begin{equation}
{\frac {d^{2}H}{d{\rho}^{2}}} 
-{\frac {dH}{d\rho}} 
-\frac{3}{4}\,H 
=C H^{-2} \left( \frac{H}{2}
  +\frac {dH}{d\rho} 
 \right)  \,.
\end{equation}
Noticing that the equation is now a fully implicit equation, we can use Abel's transformation\cite{abel}  by setting $p(H)=-2{dH}/{d\rho}$, leading to $p(H) {dp(H)}/{dH}=4 {d^{2}H}/{d{\rho}^{2}}$ so that the equation is
\begin{equation}
p \left [ H^2 \left ( \frac{dp}{dH} + 2 \right ) + 2C \right ] = 3H^3 + 2CH  \,.
\end{equation}
The immediate guess $p(H)=H$  leads to $W(r)={\rm constant}$ which, as we have seen, violates the boundary conditions.
We can go further by setting $p(H)=q(F)$ with $F=2H-2C/H$ or $H=\frac{1}{4}(F-\sqrt{F^2+16C})$ to match the flat metric to get an equation of Abel, second type, class B\cite{polyanin}:
\begin{equation}
q \left ( \frac {dq}{d{F}}+1 \right ) =
 \frac{F(3F^2 + 44C) - (3F^2 + 20C)\sqrt{F^2 +16C}}{4 ({F}^{2}-F\sqrt {{F}^{2}+16C}+16C) } \,.
\end{equation}
This has no known solution.  We also tried an inversion, such as writing the equation for $r$ as a function of $W$,
\begin{equation}
r \frac{d^2 r}{dW^2} + \left ( \frac{dr}{dW} \right )^2  \left (1 + \frac{rC}{W^2} \right ) = 0 \,.
\end{equation}
All to no avail.   Solving the equations as a power series in $1/r$, we obtain the approximate solution
\begin{eqnarray}
N &=& -1 -\frac{4M}{r} -\frac{7M^2}{r^2} - \frac{8M^3}{r^3} - \frac{8M^4 -2CM/3}{r^4} + O(r^{-5}) \,,
\nonumber \\
\alpha &=& 1 - \frac{C}{3r^3} - \frac{2CM^2}{5r^5} + O(r^{-6})\,,
\nonumber \\
\beta &=& 1 - \frac{M^2}{r^2} + \frac{2C}{3r^3} + \frac{4CM^2}{15r^5}+ O(r^{-6}) \,,
\label{Nabexpanded}
\end{eqnarray}
which agrees with Eq.\ (\ref{Nba}) when $C = 0$, but provides little help toward finding an exact solution.   

What could be the meaning of this additional class of solutions to Einstein's equations?   It is clear from both the textbook solution of Sec.\ \ref{sec2}, as well as from more rigorous considerations, that the Schwarzschild metric is unique.  Therefore, the $C \ne 0$ case cannot correspond to a physically new solution.  Accordingly, it must be related to the freedom of coordinates.  

To explore this issue, we return to the gauge condition (\ref{eq:gauge}).  By treating each of the coordinates established by harmonic gauge as a Cartesian coordinate, with the property that $\partial_\alpha x_{\rm H}^\beta = \delta_\alpha^\beta$, it is easy to show that Eq.\ (\ref{eq:gauge}) can be written in the form
\begin{equation}
\partial_\beta \gothg^{\alpha\beta}=\partial_\beta \left (\sqrt{-g} g^{\gamma\beta} \partial_\gamma  X_{\rm H}^{(\alpha)} \right ) = \Box_g X_{\rm H}^{(\alpha)} = 0 \,,
\end{equation}
where $X_{\rm H}^{(\alpha)}$ stands for the four harmonic functions $t_{\rm H}$, $x_{\rm H}$, $y_{\rm H}$ and $z_{\rm H}$, each treated as a scalar field, and $\Box_g$ is the scalar d'Alembertian in curved spacetime.   The name ``harmonic coordinate'' derives from the fact that each coordinate thus satisfies the homogeneous scalar wave equation in curved spacetime.   But because each coordinate is viewed as a scalar field, the wave equation can be expressed in any coordinate system, and in particular, in Schwarzschild coordinates.  It is easy to see that the equation for the time coordinate  $X_{\rm H}^{(0)}$ is trivially satisfied.  In a spherically symmetric geometry, the angular coordinates are clearly unaffected by this gauge choice.  Thus, if we express each Cartesian spatial harmonic coordinate as a function of the Schwarzschild coordinate $r_{\rm S}$, and the angles $\theta$ and $\phi$ according to
\begin{eqnarray}
x_{\rm H} &=& r_{\rm H}(r_{\rm S}) \sin \theta \cos \phi \,,
\nonumber \\
y_{\rm H}&=& r_{\rm H}(r_{\rm S}) \sin \theta \sin \phi \,,
\nonumber \\
z_{\rm H} &=& r_{\rm H}(r_{\rm S}) \cos \theta \,,
\end{eqnarray}
it is simple to show that each wave equation $\Box_g X_{\rm H}^{(j)} = 0$ leads to the same differential equation for $r_{\rm H}(r_{\rm S})$, namely
\begin{equation}
(r_{\rm S}^2 - 2Mr_{\rm S})r_{\rm H}'' + 2(r_{\rm S} - M)r'_{\rm H} -2r_{\rm H} = 0 \,,
\end{equation}
where $' = d/dr_{\rm S}$.  This equation is easily recognized as the Legendre equation for $\ell = 1$, with general solution $r_{\rm H} = AP_1(z) + BQ_1(z)$, where $z= (r_{\rm S}-M)/M$, and $P_1(z) =z$ and $Q_1(z) = \frac{1}{2} z \ln |(z+1)/(z-1)| -1$ are the Legendre polynomial and Legendre function, respectively.   Choosing the constant $A$ so that the two radial coordinates coincide at infinity and rescaling $B$ appropriately, we can write the final solution to the gauge condition in the form
\begin{equation}
r_{\rm H} = r_{\rm S} - M + B \left [ (r_{\rm S} - M) \ln \left (1- \frac{2M}{r_{\rm S}} \right ) + 2M \right ] \,.
\label{eq:transformation}
\end{equation}
Choosing $B=0$, we recover the relation between the harmonic and Schwarzschild radial coordinates corresponding to the case $C=0$, Eq.\ (\ref{eq:radialtransform}).  It is therefore evident that a choice $B \ne 0$ corresponds to the case $C \ne 0$.  The transcendental nature of Eq.\ (\ref{eq:transformation}) explains the difficulty of finding simple solutions for Eq.\ (\ref{eq:W}).   In the large $r$ limit, applying the expansion of the  coordinate transformation (\ref{eq:transformation}) to the Schwarzschild metric of Eq.\ (\ref{Schmetric}) yields the expanded solutions shown in Eqs.\ (\ref{Nabexpanded}) if $C = 4BM^3$.

The gauge freedom expressed in Eq.\ (\ref{eq:transformation}) should not come as a total surprise.  Students of general relativity will have encountered this already in the context of the linearized vacuum field equations.  There, Lorenz gauge leads to a flat-spacetime wave equation for the field $h^{\alpha\beta}$ (which is directly related to the linearized version of $\gothg^{\alpha\beta}$), together with the requirement that the waves be transverse, leaving 6 components of the field  $h^{\alpha\beta}$ unconstrained.  But an additional coordinate transformation can be made that maintains Lorenz gauge, as long as each of the coordinate change functions $\delta x^\alpha$ is a solution of the homogeneous flat spacetime wave equation.  Those functions allow one to constrain four more components of $h^{\alpha\beta}$, leaving only the two physically measurable modes of polarization of the gravitational wave.  

In the Schwarzschild context we see a similar phenomenon -- a class of radial coordinates that maintain harmonic gauge, yet that can change the form of the metric.   It is not clear whether the metric induced by these transformations can be expressed in any kind of simple or closed form.

\section{Discussion}
\label{sec5}

We have found that, while coordinates are irrelevant for physical quantities, the proper choice of coordinates can be critical for finding solutions of Einstein's equations. 

 It is interesting to speculate what Schwarzschild would have done had he been handed Einstein's equations only in the Landau-Lifshitz form.  Although the equations are much more complicated than the ones he dealt with, he would surely have found the relevant solution, for the $C=0$ case.  But for the case $C \ne 0$, what would he have done?  Thrown up his hands and asserted that he could not find a general exact solution to Einstein's equations?  Given the primitive understanding of general covariance of his day, would he or his contemporaries (including Einstein) have been able to recognize the additional solutions as being the same physical metric but expressed in strange coordinates?   Luckily perhaps for the history of black holes, the Landau-Lifshitz version of Einstein's equations wasn't formulated until many decades after Schwarzschild found his solution, using reasonably ``good'' coordinates.
 
Recently, Deser has presented another apparently simple derivation of the Schwarzschild metric using the Arnowitt-Deser-Misner (ADM) formalism, leading to the metric in isotropic coordinates\cite{deser}.

\begin{acknowledgments}

This work was supported in part by the National Science Foundation under grant Nos. PHY 09-65133, 12-60995, and by the Natural Sciences
and Engineering Research Council of Canada.  PF is grateful to the University of Florida for its hospitality during an internship sponsored by the \'Ecole Normale.  CMW is grateful to the Institut d'Astrophysique de Paris for its hospitality during the completion of this work.   

\end{acknowledgments}


\begin{thebibliography}{99}
\bibitem{carville}
Our title paraphrases a famous political dictum.  During Bill Clinton's first campaign for the US Presidency, his top  advisor James Carville tried to keep him focused on talking about the previous administration's economic failures by repeatedly telling him: ``It's the economy, stupid!''
\bibitem{myth}
The legend that Schwarzschild found the solution while in a trench on the battlefield (there weren't trenches at the Russian front) is evidently a myth.
\bibitem{note}
Ironically the authors of the English translation of Schwarzschild's paper cited in Ref. \cite{antoci99} state in their introduction that his paper ``leaves no room for the science fiction of the black holes''.  Evidently, confusion still reigns in some quarters.
\end{thebibliography}
\end{document}